Research Article

**Amino acid balancing for the prediction and evaluation of protein concentrations in cell-free protein synthesis systems**

Jascha Rolf[1]

Julian Handke[1]

Frank Burzinski[1]

Stephan Lütz[1]

Katrin Rosenthal[1, 2]

[1] Department of Biochemical and Chemical Engineering, Chair for Bioprocess Engineering, TU Dortmund University, Emil-Figge-Str. 66, 44227 Dortmund (Germany)

[2] School of Science, Constructor University, Campus Ring 1, 28759 Bremen (Germany).

**Correspondence:** Prof. Dr. Katrin Rosenthal (krosenthal@constructor.university). School of Science, Constructor University, Campus Ring 1, 28759 Bremen (Germany).

Dedicated to Karl-Erich Jaeger for his pioneering contributions in the field of molecular enzyme technology.

**Plain Language Summary**

Cell-free protein synthesis systems (CFPS) are an attractive tool for synthetic biology. Although many CFPS systems are described in the literature, their performance cannot be compared because the reaction components are usually used at different concentrations. Here, we present a tool based on amino acid balancing to evaluate the performance of CFPS by determining the fractional yield as the ratio between theoretically achievable and achieved protein concentration. The methods and tools can be used to quickly determine when a system has reached its maximum or is reaching other limits. The approach will facilitate the evaluation of existing CFPS systems and provide the basis for the systematic development of new CFPS systems and the optimization of existing systems.

**Abstract**

Cell-free protein synthesis (CFPS) systems are an attractive to complement the usual cell-based synthesis of proteins, especially for screening approaches. The literature describes a wide variety of CFPS systems, but their performance is difficult to compare since the reaction components are often used at different concentrations.

Therefore, we have developed a calculation tool based on amino acid balancing to evaluate the performance of CFPS by determining the fractional yield as the ratio between theoretically achievable and achieved protein molar concentration. This tool was applied to a series of experiments from our lab and to various systems described in the literature to identify systems that synthesize proteins very efficiently and those that still have potential for higher yields. The well-established *Escherichia coli* system showed a high efficiency in the utilization of amino acids, but interestingly, less-attention-paid systems, such as the one based on *Streptomyces lividans*, also demonstrated exceptional fractional yields of 100%, meaning complete amino acid conversion.

The methods and tools described here can quickly identify when a system has reached its maximum or has other limitations. We believe that the approach described here will facilitate the evaluation and optimization of existing CFPS systems and provide the basis for the systematic development of new CFPS systems.

# 1 Introduction

Cell-free protein synthesis (CFPS), also known as cell-free gene expression or in vitro protein synthesis, describes a process in which the lysate of disrupted cells or components isolated from them are used to express exogenous nucleic acids.[1] This is possible because the protein biosynthesis apparatus can act autonomously in isolation from the living cell and is still functionally present in the cell extract after cell disruption and any preparation steps.[2] CFPS thus represents the in vitro image of the living cell. It consists of a complex network of different enzymes that catalyze the synthesis of proteins starting from a gene template coding for the target protein.

A large number of different applications have already been published.[1,3] CFPS is primarily used as a complement to the otherwise common cell-based synthesis of recombinant proteins, for example, to accelerate screening of enzyme variants.[4] However, various limitations to the applicability of CFPS have also been demonstrated. When using unpurified extract-based systems, the presence of endogenous proteins or metabolic pathways can compete with the target protein.[5,6] In such a case, so-called PURE (protein synthesis using recombinant elements) systems can be used, which consist exclusively of components required for protein biosynthesis.[6] Furthermore, even though comparatively high protein concentrations up to mg per mL scale have already been achieved in CFPS systems[7], the typical yields and thus concentrations are significantly lower than in many applications using proteins obtained in vivo. This can be a problem if the intrinsic activity of the enzyme of interest is too low to be measured under assay conditions.[5] Typically, CFPS on a small-scale (<100 µL) is followed by further dilution with reaction buffer and substrates for the biocatalytic reaction. Assuming an enzyme concentration of 275 µg per mL reaction (average concentration according to NEBExpress® manual), this would mean that if the same volume of reaction buffer was added, the concentration in the assay would be only 137.5 µg per mL. For enzymes with high catalytic activity, this is a sufficient amount, but for enzymes that are not intended to catalyze their natural reaction, for example, and therefore have greatly reduced activity, such a low concentration can be problematic. Several studies have already tried to elucidate the limitation for higher protein yields and it has been suggested that different components or reaction steps could be limiting, such as translation initiation and tRNA supply[8], or an inhibitory effect of phosphate accumulation for different enzymes[9,10], such as the T7 RNA polymerase.[11] In most studies, the focus was rather on kinetic parameters or the consideration of the energy supply than on achieving higher yields.[12] A comparatively simple kinetic protein synthesis model is described which enables to predict possible yields for different PURE system compositions.[13,14] However, these studies are focused on the kinetics of the reaction and do not consider possible achievable final concentrations. To the best of our knowledge, the absolute theoretically achievable protein concentration for CFPS systems has not yet been determined. We therefore established a calculation tool based on a balancing of the building blocks for protein synthesis. Of course, translation in particular is a multisubstrate reaction that requires additional substrates besides amino acids. These are, for example, ATP, GTP, and also tRNAs. However, unlike amino acids, these molecules can be regenerated in the CFPS reaction, so no simple balancing is possible. The calculation can be used to predict the maximum achievable protein concentration based on the amino acids used and thus evaluate CFPS performance by determining the fractional yield as the ratio between theoretically achievable and actually achieved protein concentration (Equation 1).

$$\text{fractional yield} = \frac{c_{\text{Protein, experimental}}}{c_{\text{Protein, theoretical}}} \cdot 100\% \ [\text{mM}/\text{mM}] \qquad \text{Equation 1}$$

In addition, different CFPS systems can be easily compared with the use of this calculation. Many different systems based on a wide variety of origin organisms often differ in their reported maximal protein concentration. This might be due to the source organisms used for the preparation of the cell extract, but also to the composition of the buffer, especially the concentration of amino acids. Here, we present a simple method based on an amino acid mass balance to predict the maximal theoretical protein concentration of a certain protein of interest allowing a comparison independent on the source organism and amino acid concentration used.

## 2 Materials and Methods

### 2.1 Plasmids

The construction of the cyclic GMP-AMP synthase (cGAS) fusion proteins was described in Rolf et al.[5] and the construction of *Ropa*AzoRsfGFP was performed accordingly starting from pET16bP_AzoRo[15]. Plasmids were prepared using the GeneJET Plasmid Midiprep Kit (ThermoFisher Scientific, Waltham, MA, USA) or the NucleoSpin Plasmid Miniprep Kit (Macherey-Nagel, Düren, Germany), followed by a second purification with the NucleoSpin Gel and PCR Clean-up Kit (Macherey-Nagel, Düren, Germany). Elution was performed with nuclease-free water. A list of nucleotide sequences of the genes used in this study are provided in the Supplemental Material.

### 2.2 Cell-free protein synthesis

The *E. coli* extracts were prepared as described by Rolf et al.[16]. Extract-based CFPS with a reaction volume of 10 µL were performed in microtubes containing: *E. coli* cell-free extract amounting to 9.6 to 14.4 mg mL$^{-1}$ protein, 10 mM magnesium glutamate, 130 mM potassium glutamate, 1.5 or 2.5 mM each of 20 amino acids (except leucine), 1.25 or 2.25 mM leucine, 50 mM HEPES, 1.5 mM ATP and GTP, 0.9 mM CTP and UTP, 0.2 mg mL$^{-1}$ *E. coli*-tRNA, 0.26 mM CoA, 0.33 mM NAD, 0.75 mM cAMP, 0.068 mM folinic acid, 1 mM spermidine, 30 mM 3-PGA, and 2% PEG-8000. CFPS with the PUREfrex2.0 and the PURExpress system were carried out in microtubes according to the manufacturer's instructions with a reaction volume of 10 µL. All reactions were incubated in an Eppendorf® ThermoMixer® C for 4 h at 450 rpm and 37°C.

### 2.3 Quantification of protein concentration

Quantification of protein concentrations were performed based on fluorescence measurements. Two µL of the CFPS reactions were diluted in 98 µL of 0.5 M HEPES buffer (pH 8) and protein concentrations were quantified by measuring fluorescence using a FLUOstar® Omega multi-mode microplate reader (BMG LABTECH, Ortenberg, Germany) with excitation at 485 nm and emission at 520 nm in 96-well black plates. A standard curve was similarly established with purified sfGFP. Quantification was achieved by subtracting raw fluorescence signals for non-template control from the values acquired for synthesis samples and interpolating the adjusted value against the calibration curve.

## 3 Results and Discussion

### 3.1 Amino acid balance

CFPS is a biocatalytic process in which amino acids as starting materials are converted into a target protein. The protein concentrations achieved in CFPS systems is often referred to as protein yields representing the mass of protein per volume of the CFPS reaction solution. However, the more accurate definition of a yield in chemistry is as follows: the amount obtained of a final product of a reaction in relation to the amounts of substrate used.[17] Such a calculation is only possible for a component for which the exact stoichiometric relationship for the overall reaction is known. This is the case for the amino acids in CFPS systems, since it can be assumed that they are used almost exclusively as building blocks for protein synthesis. In this way, a yield related to each amino acid can be calculated. Furthermore, such a balancing of the amino acids used for protein synthesis allows the calculation of the theoretically maximum achievable protein concentration in a CFPS reaction (Equation 2).

$$c_{Protein,max} = \min \left( \frac{c_{A,CFPS}}{n_{A,Protein}}, \dots, \frac{c_{i,CFPS}}{n_{i,Protein}}, \dots, \frac{c_{V,CFPS}}{n_{V,Protein}} \right) \qquad \text{Equation 2}$$

In Equation 2, $c_{Protein,max}$ is the theoretically maximum achievable protein concentration, $c_{i,CFPS}$ is the concentration of amino acid i in the CFPS mix, $n_{i,Protein}$ is the amount of substance of amino acid i in the target protein. In order to facilitate researchers working on CFPS to determine the expected protein concentration, we have written an easy-to-use spreadsheet that, by entering the amino acid concentrations and amino acid sequence, performs the complete calculations (available in the Supplemental Material). In addition to the maximum possible protein concentration, actually achieved concentrations can be entered, so that the yield related to each amino acid and fractional yield as the ratio between theoretically achievable and actually achieved protein concentration (Equation 1) can be determined. This allows an assessment of how well the synthesis for the target protein proceeds in the CFPS system used. In addition to the spreadsheet, we have also programmed a Jupyter notebook so that, depending on application preference, this pre-programmed tool can also be used to determine the maximum possible protein concentration and fractional yield (available on request). Figure 1 shows the required inputs (amino acid sequence, amino acid concentrations used, and protein concentration obtained; boxed in green), as well as the corresponding output provided by the Jupyter Notebook. Furthermore, the program allows to easily identify actual limiting amino acids.

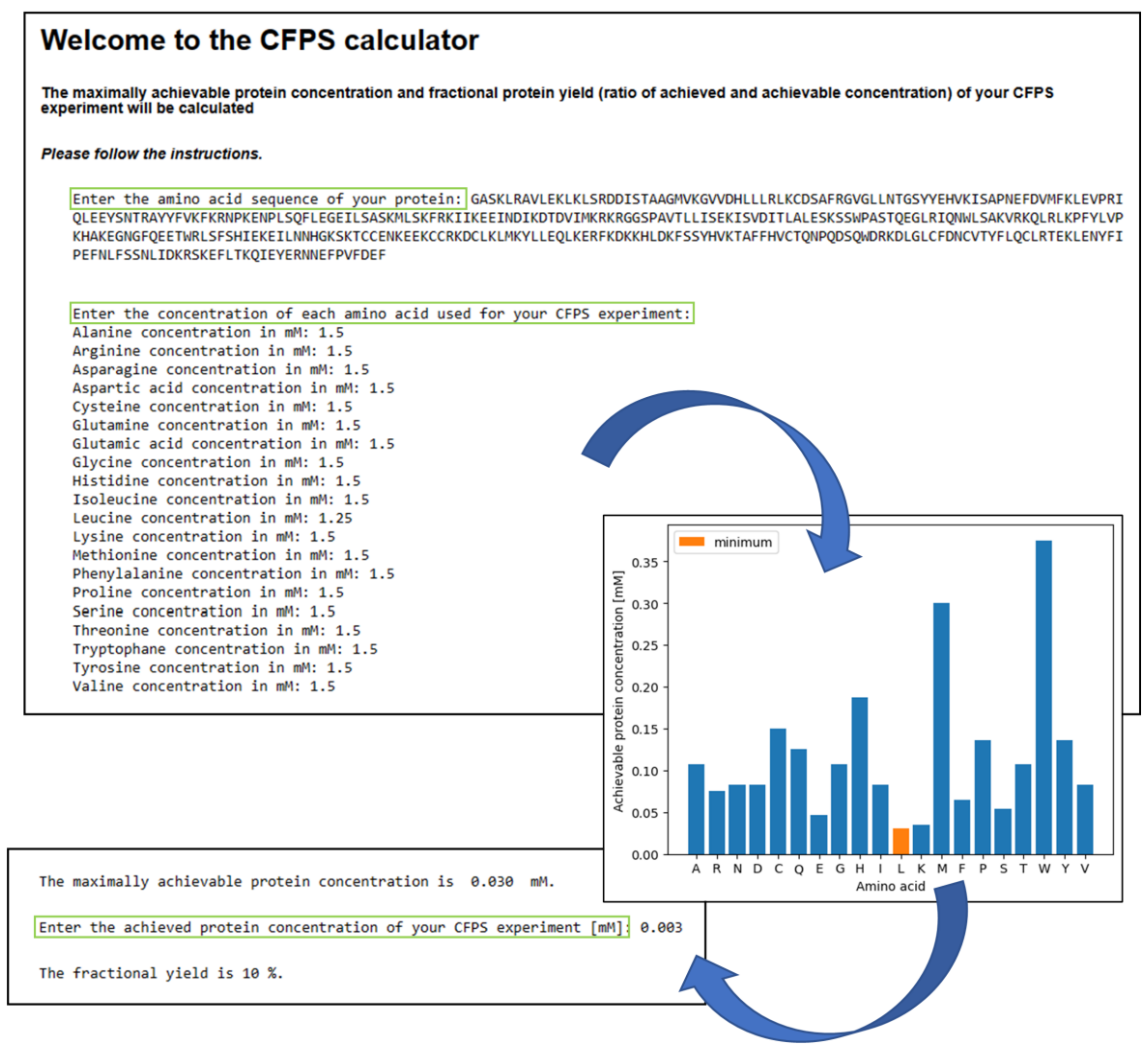

Figure 1. Jupyter Notebook interface of the CFPS calculator.

In systems based on non-purified extracts, additional residual amino acids are introduced by the extract because the cells are usually processed in the mid-exponential growth phase and grow on complex media that are rich in nutrients. To evaluate whether these residual amino acids originating from the intracellular composition of the cells have a significant influence on the calculations, we have estimated the additional amounts of amino acids to be expected. For this purpose, we have carried out the estimation based on our in-house CFPS system. Assuming a cell harvest at an $OD_{600}$ of 3 and a concentration factor of 10 for the extract processing steps, the theoretically obtained $OD_{600}$ of the extract is 30. Using a factor of 0.312 to convert to gram cell dry weight per liter (determined in our laboratory), the corresponding cell dry weight used for the preparation of 1 L of the cell extract is 9.36 g. Assuming the intracellular concentration of 11 proteinogenic amino acids[18], the amino acid concentrations thus range from 2 µM for phenylalanine to 1 mM for glutamate in the cell-free extract. With an extract fraction of about 25% in the CFPS reaction, 0.5 µM to 15 µM amino acids are introduced into the system by the extract. Only glutamate is an exception here with 235 µM. In the case of glutamate, however, it should be noted that glutamate is also present in many CFPS system buffers, so that the concentration of glutamate in the reaction is above 100 mM. Thus, with a usual use of 1.5 mM of each amino acid, no more than 1% additional amino acids are introduced by the extract, with the exception of glutamate. The release of intracellular amino acids was therefore neglected in the calculation.

### 3.2 Evaluation of CFPS performance for different proteins

We used the tool to calculate the theoretical maximally possible concentrations as well as the fractional yield for the synthesis of various proteins produced in our laboratory (Table 1 & Figure 2).

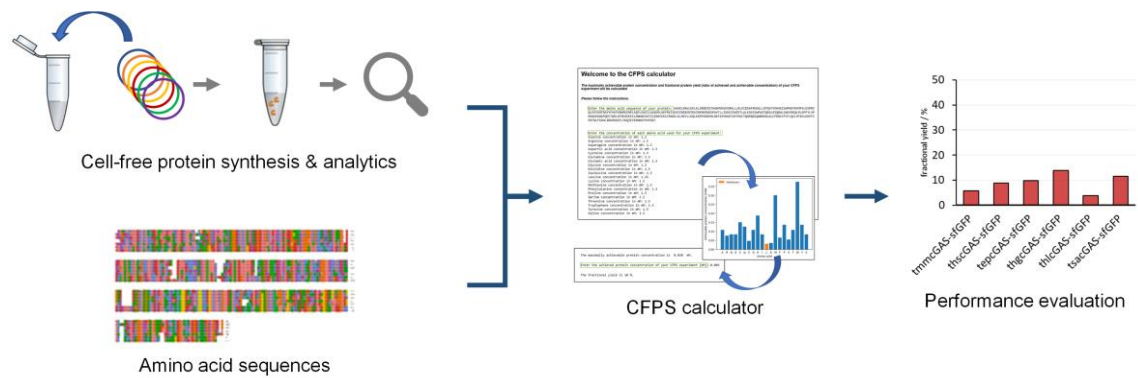

Figure 2. Workflow to evaluate the performance of cell-free protein synthesis for different enzymes.

Table 1. Calculated maximum protein concentrations based on the quantity of amino acids used, actual protein concentrations obtained in cell-free protein syntheses and the corresponding fractional yield.

| Protein | Amino acid concentration [mM] | Maximum protein concentration [mg mL$^{-1}$] | Actual protein concentration [mg mL$^{-1}$] | Fractional yield [%] |
|---|---|---|---|---|
| **In-house CFPS** | | | | |
| sfGFP | 1.5 (1.25 leucine) | 1.65 | 0.96 ± 0.04 | 58.2 ± 3.1 |
|  | 2.5 (2.25 leucine) | 2.90 | 1.20 ± 0.00 | 41.4 ± 0.0 |
| t*mm*cGAS-sfGFP | 1.5 (1.25 leucine) | 1.52 | 0.09 ± 0.01 | 5.7 ± 0.6 |
| + chaperones | 1.5 (1.25 leucine) | 1.52 | 0.13 ± 0.01 | 8.5 ± 0.7 |
| t*hs*cGAS-sfGFP | 1.5 (1.25 leucine) | 1.49 | 0.13 ± 0.02 | 8.9 ± 1.5 |
| t*ep*cGAS-sfGFP | 1.5 (1.25 leucine) | 1.62 | 0.16 ± 0.01 | 9.8 ± 0.5 |
| t*hg*cGAS-sfGFP | 1.5 (1.25 leucine) | 1.53 | 0.21 ± 0.01 | 13.8 ± 0.7 |
| t*hl*cGAS-sfGFP | 1.5 (1.25 leucine) | 1.42 | 0.05 ± 0.00 | 3.7 ± 0.2 |
| t*sa*cGAS-sfGFP | 1.5 (1.25 leucine) | 1.70 | 0.20 ± 0.01 | 11.6 ± 0.8 |
| **PUREfrex2.0** | | | | |
| sfGFP | 0.3 | 0.35 | 0.24 ± 0.07 | 67.7 ± 19.5 |
| *Ropa*AzoR-sfGFP | 0.3 | 0.32 | 0.05 ± 0.01 | 14.2 ± 3.3 |
| **PURExpress** | | | | |
| sfGFP | 0.3 | 0.35 | 0.15 ± 0.07 | 43.6 ± 18.8 |

sfGFP: superfolder green fluorescent protein; t*mm*cGAS: truncated *Mus musculus* cyclic GMP-AMP synthase; t*hs*cGAS: truncated *Homo sapiens* cyclic GMP-AMP synthase; t*ep*cGAS: *Equus przewalskii* cyclic GMP-AMP synthase; t*hg*cGAS: *Heterocephalus glaber* cyclic GMP-AMP synthase; t*hl*cGAS: *Haliaeetus leucocephalus* cyclic GMP-AMP synthase; t*sa*cGAS: *Sinocyclocheilus anshuiensis* cyclic GMP-AMP synthase; *Ropa*AzoR: *Rhodococcus opacus* azoreductase

In each case, the theoretical maximum protein concentrations were in a similar range for all proteins synthesized with the in-house CFPS, 1.5-1.7 mg mL$^{-1}$, as well as for the proteins synthesized with different PURE systems, around 0.35 mg mL$^{-1}$. The difference between the systems can be explained by the provision of different amino acid concentrations; in the in-house system[5], 1.5 mM of each amino acid was added (except of leucine), while the amino acid concentration in the PURE system was 0.3 mM.[19] Leucine was used at lower concentrations of about 1.25 mM in the in-house system due to practical reasons. Leucine is poorly soluble in water and therefore a lower concentrated stock solution was used for this amino acid, which resulted in a lower concentration in the reaction mixture.

Interestingly, the experimentally achieved protein concentration was considerably lower than the theoretical maximum possible concentrations. The obtained protein concentrations were between 0.05-1.20 mg mL$^{-1}$ and seemed not to be dependent on the theoretically achievable protein concentration, but to be determined by the amino acid sequence of the protein. The homologous proteins, which encode the cyclic-GMP-AMP synthases (cGASs) fused with a super folder GFP variant (sfGFP), all showed low protein concentrations, reaching only between 4 and 14% of the theoretically possible protein concentrations based on the amino acids used. This suggests that some other component or reaction step of protein synthesis is limiting. The situation is different for the synthesis of sfGFP alone, which is one of the standard proteins for determining the synthesis performance of CFPS systems. Both the in-house system and the PURE systems achieved fractional yields of 41 to 68% for the synthesis of sfGFP.

To determine how increased amino acid concentrations affects the fractional yield, the concentrations were increased to 2.5 mM (leucine 2.25 mM) for the in-house CFPS system. Interestingly, although a higher protein concentration of 1.2 mg mL$^{-1}$ could be achieved, the fractional yield decreased to 41% (Table 1). Even though this corresponds to one of the highest protein concentrations achieved in a CFPS system, it can be seen that the amino acids provided are utilized less efficiently by the system. This is a relevant information when considering the costs of CFPS systems. Amino acids are often the components with the largest amount in the system and therefore account for about 8% of the cost in *E. coli*-based CFPS systems, for example.[20] Therefore, the calculation of fractional yield provides a new target value for the optimization of CFPS systems. The PURE systems contain significantly lower amino acid concentrations than most extract-based *E. coli* systems. This simply explains the often generally lower protein concentrations achieved. However, the highest fractional yield in our experiments was achieved with PUREfrex2.0. This indicates that the system is highly optimized, and the reactants used are efficiently converted into the product. The influence of the use of specially adapted extracts can also be considered in more detail. For example, we have developed a chaperone-enriched extract for the synthesis of difficult-to-synthesize enzymes[16] and have also tested it for the synthesis of cGAS (Table 1). In this case, better synthesis performance is also accompanied by better utilization of amino acids. Since the presented tool only requires information about the amino acid concentration of a system, it can also be easily applied to all non-*E. coli* CFPS systems. Especially when developing new systems, it can be useful to carry out such an efficiency analysis.

**3.3 Evaluation of CFPS systems originating from different source organisms**

In a next step, we used the calculation tool for the comparison of different CFPS systems described in the literature. CFPS systems based on different prokaryotic and eukaryotic organisms used in batch processes were considered. Based on the amount of amino acids in the system and the amino acid sequence of the reporter proteins, the fractional yields were calculated (Figure *3*).

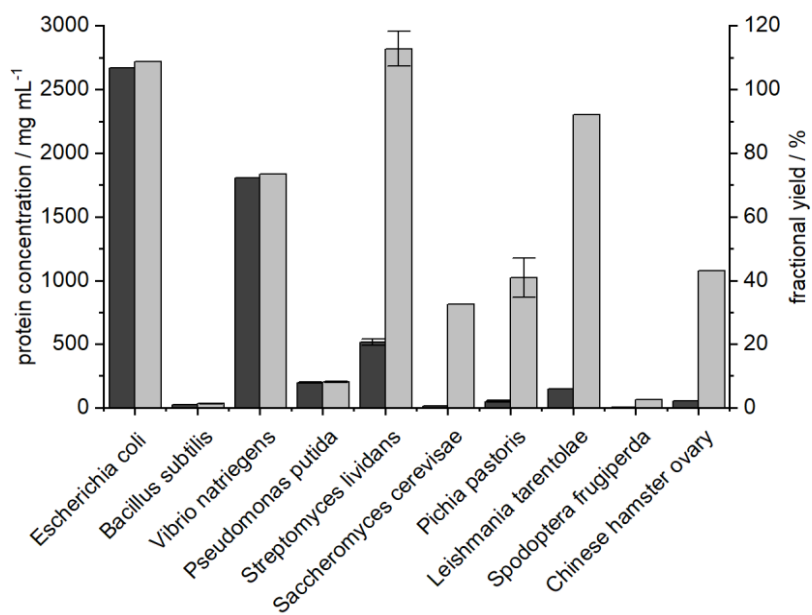

Figure 3. Achieved protein concentrations (dark grey) and fractional yield (light grey) of cell-free protein synthesis systems starting from different organisms.

The protein concentrations achieved differ greatly between the different systems, but also the amount of amino acids used varies considerably (Table *2*).

Table 2. Comparison of different cell-free protein synthesis systems based on achieved protein concentrations and their fractional yield considering the quantity of used amino acids.

| Organism | Protein | Molecular weight [kDa] | Protein concentration [µg mL$^{-1}$] | Amino acid concentration [µM] | Fractional yield [%] | Reference |
|---|---|---|---|---|---|---|
| *Escherichia coli* | sfGFP[a] | 27.0 | 2670.0 ± 0.1 | 2000 | 108.7 ± 0.0 | [21] |
| *Bacillus subtilis* | GFPmut3b[b] | 26.9 | 22.8 ± 1.2 | 1500 (Leu 1250) | 1.3 ± 0.1 | [22] |
| *Vibrio natriegens* | sfGFP[a] | 27.0 | 1804.0 | 2000 | 73.5 | [23] |
| *Pseudomonas putida* | sfGFP[a] | 26.8 | 198.0 ± 5.9 | 2000 | 8.1 ± 0.2 | [24] |
| *Streptomyces lividans* | eGFP[c] | 26.9 | 515.7 ± 25.3 | 373 | 112.8 ± 5.5 | [25] |
| *Streptomyces venezuelae* | mScarlet-I[d] | 28.1 | 266 | 1500 (Leu 1250) | 17.1 | [26] |
| *Saccharomyces cerevisae* | sfGFP[a] | 27.0 | 16.0 | 40 | 32.6 | [27] |
| *Pichia pastoris* | sfGFP[a] | 27.0 | 50.2 ± 7.5 | 100 | 40.9 ± 6.1 | [28] |
| *Leishmania tarentolae* | eGFP[c] | 26.9 | 150.0 | 133 | 92.1 | [29] |
| *Spodoptera frugiperda* | Luciferase | 60.6 | 6.1 | 200 | 2.6 | [30] |
| Chinese hamster ovary | Luciferase | 60.6 | 51.3 | 100 | 43.1 | [31] |

*a* superfolder green fluorescent protein *b* optimized mutant of green fluorescent protein *c* enhanced green fluorescent protein *d* red fluorescent protein

Thus, an actual comparison about the performance of the systems is not possible simply based on the protein concentrations achieved. The fractional yields were therefore calculated to identify systems where there is still plenty of potential for improvement. The systems based on *Bacillus subtilis*[22], *Pseudomonas putida*[24], and *Spodoptera frugiperda*[30], produced less than 10% of the theoretically achievable protein amounts. On the one hand, this may indicate another limiting factor in the system, and on the other hand, amino acids may also be consumed in non-intended reactions, limiting the yield of synthesized products. Other systems, such as the *E. coli* system, which is probably the most researched and optimized system, already can achieve full utilization of the amino acids provided.[21] The *Streptomyces lividans* system also completely converts the added amino acids.[25] Actual fractional yields above 100% can be explained by the standard deviation and evaporation effects over the synthesis time. In addition, the *Vibrio natriegens*-based CFPS system achieves very high yields in relation to the substrates used.[23] Interestingly, the second *Streptomyces* system considered in our calculations showed lower protein concentrations using larger amounts of amino acids resulting in significantly lower fractional yields.[26] However, the concentration of the reporter protein could be significantly increased for this system using a novel approach.[32] Instead of a commercial standard, rich media components were added as amino acid sources. Especially tryptone and casamino acids gave up to a 2-fold increase in protein concentration for the *S. venezuelae* system. This indicates that amino acid availability plays a limiting role for this system. The use of rich media thus also represents an interesting and cost-effective alternative to the defined addition of amino acid standards.

Some of the systems described are still unoptimized and may also be limited by synthesis time, for example, but it is the approach described here that makes it easy to quickly assess how far one is from theoretically possible concentrations. Very high fractional yields give an indication that indeed the amino acid concentration may be the limiting component for even higher protein concentrations. However, it has been already shown that different CFPS systems respond differently to changes in the system`s reagent component composition and optimization for the individual case may be necessary.[33] For this optimization, not only the synthesis system under consideration can play a role, but also the design of the DNA template and the target protein itself. For example, it has been shown that the choice of the expression vector plays a critical role in CFPS, as vectors may contain elements that negatively affect the protein expression.[34] To date, the highest protein concentration achieved in batch-mode and described in literature is about 4 mg mL$^{-1}$ of deGFP and was synthesized by an *E. coli*-based system.[35] The applied amino acid concentrations differed in this synthesis depending on the amino acid in a range of 1.5 to 3 mM. Assuming that the most abundant amino acid of the protein, which is glycine (9.5%) was used at 3 mM, a fractional yield of approximately 100% was achieved in this approach. But not only *E. coli* systems can reach such high concentrations. Also, a eukaryotic system based on *Nicotiana tabacum* reaches protein concentrations of up to 3 mg mL$^{-1}$ and is commercialized as ALiCE® (Almost Living Cell-free Expression).[36] The peculiarity of this system is that it contains native, actively translocating microsomal vesicles derived from the endoplasmic reticulum and the Golgi and shows good capabilities of performing post-translational modifications. For this system, however, no statements can be made about any amino acids that may have been used.

**4 Concluding remarks**

Balancing the substrates and products in a reaction can quickly and easily give important insight into possible potentials and limitations. This is also true for CFPS and can be done by balancing the amino acids, which are the main substrates of synthesis. By applying the method described here and using the tool presented, it becomes very quickly clear whether a system is already at its maximum or whether it is otherwise limited. What constitutes such a limitation can then be investigated on a case-by-case basis, considering the source organism and the target protein. Also, a comparison of the multitude of systems described in the literature only becomes possible when not only the maximum protein concentration achieved is considered, but also the amount of substrate used. In this way, systems that would otherwise have attracted little attention due to the low concentrations become significantly more interesting. In addition, detailed consideration of amino acid usage can help optimized systems use fewer unnecessary amino acids, leading to cost reductions. We believe that the approach described here will make it easier to evaluate existing CFPS systems and provide the basis for systematic development of new CFPS systems.

The authors have declared no conflicts of interest.

The data that supports the findings of this study are available in the supplemental material of this article. The Jupyter Notebook is available on request.

**Acknowledgements**

We thank Stephan Malzacher for discussing the development of the Jupyter Notebook Application. The plasmid pET16bP_ AzoRo was kindly provided by Dirk Tischler.